%% file: main.tex
\documentclass[11pt]{article}

\usepackage[letterpaper,margin=1in]{geometry}
\usepackage[utf8]{inputenc}
\usepackage[T1]{fontenc}
\usepackage{times}
\usepackage[numbers,sort&compress]{natbib}
\usepackage{amsmath,amssymb}
\usepackage{booktabs}
\usepackage{graphicx}
\usepackage{float}
\usepackage{microtype}
\usepackage{xcolor}
\usepackage{hyperref}
\hypersetup{breaklinks=true}
\usepackage{url}
\usepackage{nicefrac}
\usepackage{enumitem}

\newcommand{\fge}{\textsc{FGE}}
\input{data_macros}

\title{From Forecasting to Feedback: Evaluating Learned Dynamics for Tokamak Control}
\author{
  Guoyang Shi$^{1,2,3}$, Siqi Ding$^{1,2,3}$, Xuanhe Wang$^{1,2,3}$, Pei Guo$^{4}$, Changquan Yu$^{5}$, Jiale Hu$^{6}$,\\
  Jiayi Zhi$^{7}$, Zhengyuan Chen$^{1,2,3}$, Jianguo Chen$^{1,2,3}$, Tianyuan Liu$^{1,2,3,*}$
}
\date{}

\begin{document}

\maketitle

\begin{center}
\small
$^1$Beijing ENN Fusion Energy Science and Technology Co., Ltd., Beijing 101111, China\\
$^2$Beijing Key Laboratory of High Magnetic Field Spherical Torus Fusion Energy, Beijing 101111, China\\
$^3$Hebei Key Laboratory of Compact Fusion, Langfang 065000, China\\
$^4$Dalian University of Technology, Dalian 116024, China\\
$^5$East China University of Technology, Nanchang 330013, China\\
$^6$Anhui University, Hefei 230601, China\\
$^7$Nankai University, Tianjin 300071, China\\
$^*$Corresponding author: \href{mailto:liutianyuan@enn.cn}{liutianyuan@enn.cn}
\end{center}

\input{sections/abstract}
\input{sections/introduction}
\input{sections/method}
\input{sections/results}
\input{sections/conclusion}

\bibliographystyle{IEEEtran_doi}
\bibliography{references}

\clearpage
\appendix
\input{sections/appendix}

\end{document}

%% file: data_macros.tex
\newcommand{\EvalSteps}{400}
\newcommand{\EvalSeconds}{0.4}
\newcommand{\ControlHz}{1000}

\newcommand{\TrialsPerCell}{6}
\newcommand{\NumTemporalModels}{16}
\newcommand{\NumPolicies}{8}
\newcommand{\SelectedScale}{5k}

\newcommand{\DTwoSurv}{83.3}

\newcommand{\DTwoHorizon}{99.0}

\newcommand{\DTwoMinHorizon}{94.2}

\newcommand{\DThreeHorizon}{94.7}

\newcommand{\DThreeMinHorizon}{68.2}

\newcommand{\DFourHorizon}{90.5}

\newcommand{\DFourMinHorizon}{43.2}

\newcommand{\DOneOwnErr}{0.135}

\newcommand{\DFourCommonErr}{0.381}
\newcommand{\DFourSmallCommonErr}{0.528}

\newcommand{\DFourScaleReduction}{28}
\newcommand{\DOneSmallOwnErr}{0.345}
\newcommand{\DOneSmallOODErr}{0.960}
\newcommand{\DOneOODErr}{1.060}
\newcommand{\DOneOODGap}{0.925}

\newcommand{\OverallTrials}{24}
\newcommand{\OverallNominalOK}{19}
\newcommand{\OverallNoiseOK}{5}
\newcommand{\OverallNominalSurv}{79.2}
\newcommand{\OverallNoiseSurv}{20.8}
\newcommand{\NominalOnlyPairs}{14}
\newcommand{\NoiseOnlyPairs}{0}

\newcommand{\NominalMeanSteps}{372.8}
\newcommand{\NoiseMeanSteps}{207.5}

\newcommand{\CommonSurvivorPairs}{5}

%% file: sections/abstract.tex
\begin{abstract}
Learned temporal models are moving beyond offline forecasting toward interactive environments for decision making, yet logged-trajectory accuracy need not predict closed-loop behavior under policies optimized in the learned model.  This forecasting--control gap matters in tokamak magnetic control, where coupled plasma--circuit dynamics and actuator constraints can amplify errors through feedback.  Using a calibrated free-boundary Grad--Shafranov solver, we collect four datasets in a $2\times2$ design over physics-parameter and current-reference coverage.  At matched budgets of 500, 1k, 3k, and 5k rollouts, we train \NumTemporalModels{} action-conditioned Transformers; PPO policies optimized in the 5k surrogates are then evaluated without adaptation on six reference-solver conditions.  Only joint-coverage D4 shows monotonically decreasing cross-condition mean error, whereas D1 improves in distribution but degrades out of distribution.  Binary completion masks substantial closed-loop differences: D2--D4 each complete \DTwoSurv\% of conditions, but reach on average \DTwoHorizon\%, \DThreeHorizon\%, and \DFourHorizon\% of the evaluation horizon, while their worst conditions reach \DTwoMinHorizon\%, \DThreeMinHorizon\%, and \DFourMinHorizon\%.  Policies trained with action noise complete \OverallNoiseOK{}/\OverallTrials{} matched conditions, compared with \OverallNominalOK{}/\OverallTrials{} under nominal training, and have a lower mean horizon-capped episode length.  Based on one trained policy per configuration and one evaluation realization per condition, the results support combining prediction error, episode completion, condition-level duration, and tracking quality.

\end{abstract}

%% file: sections/introduction.tex
\section{Introduction}

Magnetic control regulates plasma current, position, and shape under coupled dynamics and actuator constraints.  Reinforcement learning (RL) can optimize these objectives jointly, and simulator-trained policies have been deployed on the TCV tokamak without adaptation \mbox{\citep{degrave2022magnetic,tracey2024practical}}.  Such deployment requires a simulator that supports large-scale interaction while remaining accurate over the state--action distribution induced by the optimized policy.  Learned dynamics models fitted to archived discharges offer a scalable simulation alternative; studies on DIII-D and HL-3 demonstrate their potential for offline policy learning and experimental control \mbox{\citep{char2023offline,wu2025highfidelity,wang2025predictfirst}}.  Yet accurate prediction of logged trajectories may not imply reliable dynamics for policy optimization.

\begin{figure}[t]
  \centering
  \includegraphics[width=0.93\linewidth]{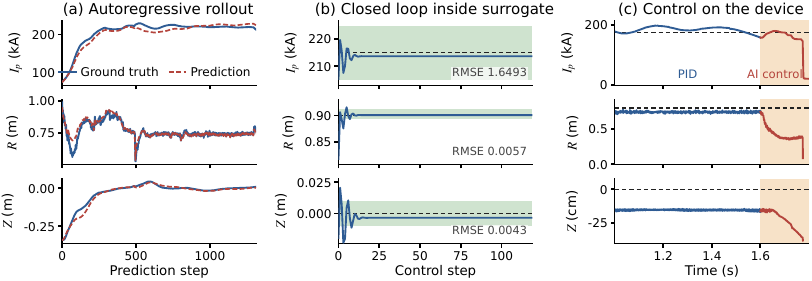}
  \caption{\textbf{Forecasting--control mismatch.}  (a) A temporal model tracks a 1,314-step rollout.  (b) Surrogate control remains within tolerance.  (c) Shot \#10311 drifts after switching from PID (blue) to the learned controller (red, originally labeled ``AI'') at 1.6\,s.  These prior observations motivate but are not included in our controlled comparison.}
  \label{fig:motivation}
\end{figure}

Open-loop prediction and closed-loop policy evaluation operate under different trajectory distributions.  Supervised validation follows trajectories generated by a behavior controller.  After RL optimization, the learned model and optimized policy jointly determine the resulting state--action distribution.  Limited coverage can permit model exploitation; state errors then alter feedback actions and compound through autoregressive rollout.  This links exposure bias in sequence modeling \mbox{\citep{bengio2015scheduled}} to distributional uncertainty in model-based and offline RL \mbox{\citep{chua2018pets,janner2019mbpo,yu2020mopo,kidambi2020morel}}.  In tokamak control, a short sequence of biased actions can rapidly degrade tracking and may eventually trigger termination; \mbox{Fig.~\ref{fig:motivation}} illustrates the initial drift.  Offline RL shows that dataset composition matters beyond sample count \mbox{\citep{fu2020d4rl}}, while dynamics randomization broadens operating distributions \mbox{\citep{peng2018dynamics}}.  For learned tokamak simulators, however, the roles of physical-parameter and reference-trajectory coverage remain unclear when data volume and collection behavior are controlled.

We address this gap using the calibrated \fge{} solver as a deterministic, instrumented reference.  We examine how data-coverage choices are associated with the reference-solver performance of policies optimized in learned surrogates.  Our evaluation separates held-out autoregressive prediction from frozen-policy completion, duration, and tracking, while D1--D4 vary physical-parameter and current-reference coverage under matched collectors and rollout budgets.  Among the evaluated dataset--model configurations, only joint-coverage D4 shows consistent cross-condition gains with scale.  Completion rate alone ties D2--D4 and obscures differences in time to termination; the tested policies trained with action noise also terminate earlier on matched conditions.  These observations motivate reporting cross-condition prediction together with multiple closed-loop reference-solver metrics.

%% file: sections/method.tex
\section{Experimental design and evaluation}
\label{sec:method}

\begin{figure}[t]
  \centering
  \includegraphics[width=\linewidth]{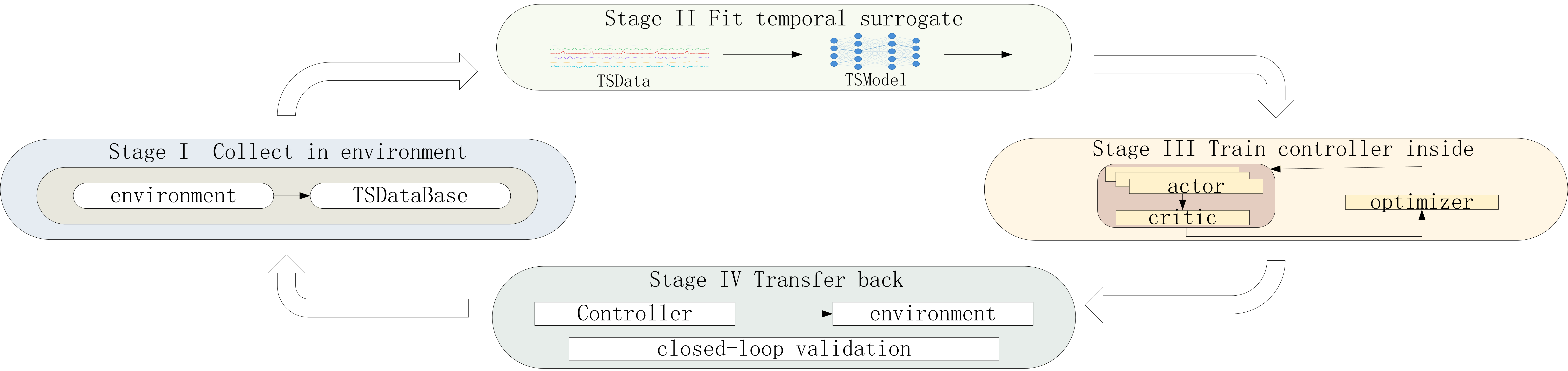}
  \caption{\textbf{Experimental workflow.}  Data are collected in the reference environment, used to fit a temporal surrogate, and then used for controller training inside the surrogate.  The trained controller is finally transferred back to the reference environment for closed-loop validation.}
  \label{fig:method-overview}
\end{figure}

\paragraph{Reference environment.}
\fge{} is a fast, time-evolving free-boundary Grad--Shafranov solver coupling plasma equilibrium, circuit dynamics, and current-profile evolution \mbox{\citep{heiss2026fge}}.  We use the configuration calibrated to EXL-50U shot 13844 \mbox{\citep{shi2025strategy,shi2025onema,shi2026overview}} as a deterministic, resettable reference for both data collection and frozen-policy evaluation.  This is a \emph{surrogate-to-reference-solver} study, not a hardware deployment.

\paragraph{Data-coverage conditions.}
D1--D4 form a $2\times2$ factorial design over physics-parameter coverage and current-reference coverage.  The same frozen RL collector runs in \fge{} at 1\,ms resolution, with the same colored action-noise process added before voltage clipping and rate limiting.  Physics coverage uses Latin hypercube sampling of multiplicative scales for $b_p$, $q_0$, $\sigma_{\mathrm{neo}}$, and $L_p$ over $[0.70,1.30]$, $[0.80,1.30]$, $[0.70,1.30]$, and $[0.80,1.00]$, respectively; reference coverage ramps $I_p$ from 495\,kA to an endpoint sampled in $[200,700]$\,kA.

\begin{table}[t]
  \caption{$2\times2$ data-coverage design.  All datasets use the same collector and action-noise process, with 500/1k/3k/5k matched trajectory counts.  C1--C4 mirror D1--D4.}
  \label{tab:data}
  \centering
  \footnotesize
  \begin{tabular}{lcccl}
    \toprule
    Set & Collection excitation & Physics LHS & $I_p$ ramp & Purpose \\
    \midrule
    D1 & \checkmark &            &            & narrow baseline \\
    D2 & \checkmark & \checkmark &            & physics coverage \\
    D3 & \checkmark &            & \checkmark & reference coverage \\
    D4 & \checkmark & \checkmark & \checkmark & joint coverage \\
    \bottomrule
  \end{tabular}
\end{table}

Collector, excitation, and rollout count within each budget are matched to control for collection behavior and data volume.  The design matches trajectory count, not effective sample diversity: one broad-coverage rollout may contain more diverse transitions.

\paragraph{Action-conditioned temporal surrogate.}
The surrogate state comprises 12 circuit-current channels, $I_p$, $(R,Z)$, boundary radii $(R_{\min},R_{\max})$, elongation $\kappa$, and $r_c$; the action is the 12-dimensional coil-voltage vector actually applied after clipping and rate limiting.  Complete rollouts, rather than overlapping windows, are assigned to disjoint training and validation partitions; C1--C4 are held out separately.  Given $H=30$ standardized past steps, the model predicts
\begin{equation}
  \hat{s}_{t+1} =
  f_\theta\!\left(\hat{s}_{t-H+1:t},\, a_{t-H+2:t+1},\, \phi\right),
  \label{eq:model}
\end{equation}
where $\phi$ supplies physics parameters for D2 and D4 and is omitted for D1 and D3.  Each dataset trains one model at 500/1k/3k/5k rollouts, giving \NumTemporalModels{} surrogates: D1 and D3 use the base version, whereas D2 and D4 use the parameter-conditioned version.  All models share the same three-layer Transformer backbone ($d_{\mathrm{model}}=128$, eight heads) \citep{vaswani2017attention} and are evaluated autoregressively on C1--C4.  Per-channel RMSEs are normalized by channel scale and equally weighted; predicted states are recursively fed back during autoregressive rollout.  Policy training uses the \SelectedScale{} checkpoints.

\paragraph{Policy optimization and reference-solver evaluation.}
PPO \citep{schulman2017ppo} observes $(I_p, R_{\min}, R_{\max}, \kappa)$ and outputs seven independent voltage commands, expanded by a fixed coil-pair mapping to 12 bounded channels.  D1--D4 use the same PPO implementation, network, hyperparameters, and training budget; learned and reference environments share reward, target generation, action bounds, and termination.  Each surrogate trains a \emph{nominal} policy without added perturbations and a \emph{noise} policy with action perturbations during PPO training; this training noise is distinct from collection excitation.  This gives \NumPolicies{} frozen controllers.

\paragraph{Closed-loop evaluation.}
The task ramps $I_p$ from 495\,kA to 400/500/600\,kA while holding boundary shape.  Each controller is evaluated at all three targets under nominal and perturbed physics.  Controllers reuse the same random seeds, preserving the randomized parameter samples for paired evaluation.  Each (policy, training regime) combination contains \TrialsPerCell{} conditions run at \ControlHz{}\,Hz for \EvalSteps{} steps (\EvalSeconds{}\,s).  We report episode-completion rate together with each condition's duration and the mean horizon reached, $\frac{1}{N}\sum_i\min(T_i,\EvalSteps)/\EvalSteps$, to distinguish failures occurring at different times.  Early termination counts as failure.  Because tracking errors from failed episodes of different lengths are not directly comparable, we report tracking errors among completed episodes only, together with the number completed, and keep the channels separate ($I_p$ in kA, radii in cm, $\kappa$ dimensionless).

%% file: sections/results.tex
\section{Results}
\label{sec:results}

\begin{table}[t]
  \caption{Reference-solver evaluation.  Policies trained in each surrogate are evaluated without adaptation in \fge{}.  Mean horizon is the average fraction of \EvalSteps{} steps reached; tracking errors include completed episodes only.}
  \label{tab:main}
  \centering
  \footnotesize
  \resizebox{\linewidth}{!}{\input{tables/main_control_en}}
\end{table}

\begin{figure}[t]
  \centering
  \includegraphics[width=\linewidth]{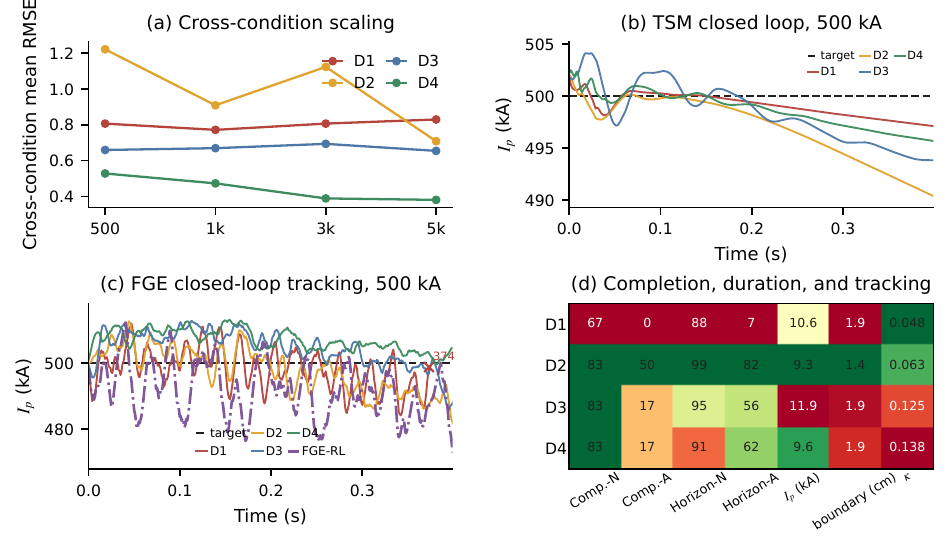}
  \caption{\textbf{Prediction and control.}  (a) Cross-condition scaling.  (b) TSM closed-loop $I_p$ at a fixed 500-kA target over 400 steps.  (c) Transferred D1--D4 policies and the directly trained FGE-RL baseline at 500\,kA without test-time physics perturbation; the cross marks termination.  (d) Completion, mean horizon, and completed-episode tracking error.  N/A denote nominal/action-noise training; colors are normalized by column with green better.}
  \label{fig:results}
\end{figure}

\paragraph{Only the joint-coverage configuration scales consistently across conditions.}
Among the evaluated dataset--model configurations, D4 is the only one for which cross-condition mean error decreases at every scale, from \DFourSmallCommonErr{} at 500 rollouts to \DFourCommonErr{} at 5k (\DFourScaleReduction\%).  D1 shows a contrasting in- versus out-of-distribution pattern: its in-distribution error falls from \DOneSmallOwnErr{} to \DOneOwnErr{}, while its out-of-distribution mean rises from \DOneSmallOODErr{} to \DOneOODErr{}, leaving a \DOneOODGap{} gap at 5k.  The 5k matrix further shows that each configuration attains its lowest error on the matching test condition (Appendix~\ref{fig:app-openloop-matrix}).  Because D1/D3 use the base model and D2/D4 the parameter-conditioned model, these comparisons do not identify the causal effect of coverage alone.

\paragraph{Surrogate-internal tracking does not determine reference-solver endurance.}
All four direct TSM loops complete 400 steps, but their long-horizon behavior differs: after the initial transient, all traces develop negative drift, with D2 and D3 departing most strongly from the fixed 500-kA target (Fig.~\ref{fig:results}b).  These surrogate-internal trajectories are separate from the six-condition FGE transfer.  At 5k, D2--D4 all reach \DTwoSurv\% completion in FGE, yet their mean horizon fractions are \DTwoHorizon\%, \DThreeHorizon\%, and \DFourHorizon\%, and their worst-condition fractions are \DTwoMinHorizon\%, \DThreeMinHorizon\%, and \DFourMinHorizon\%.  In the nominal 500-kA cell shown in Fig.~\ref{fig:results}c, D2--D4 and the directly trained FGE-RL baseline complete 400 steps, whereas D1 terminates at step 374.  FGE-RL completes the episode but exhibits substantially larger oscillations, further showing that completion alone does not imply high-quality control.

\paragraph{Policies trained with action noise terminate earlier in this evaluation.}
Across the \OverallTrials{} matched configuration--condition pairs, nominal-training policies complete \OverallNominalOK{} evaluations (\OverallNominalSurv\%) and action-noise policies complete \OverallNoiseOK{} (\OverallNoiseSurv\%; Fig.~\ref{fig:results}d).  Relative to nominal training, action-noise training is associated with \NominalOnlyPairs{} lost completions and \NoiseOnlyPairs{} additional completions; the mean horizon-capped episode length falls from \NominalMeanSteps{} to \NoiseMeanSteps{} steps.  On the \CommonSurvivorPairs{} pairs completed under both regimes, the observed $I_p$ and boundary errors are also numerically larger with action-noise training (Appendix~\ref{tab:app-paired-tracking}).  These comparisons, based on one trained policy per configuration and one evaluation realization per condition, describe the tested noise setting rather than its population-level training effect.

%% file: tables/main_control_en.tex
\begin{tabular}{lccccccc}
  \toprule
  Policy & Open-loop err.\ $\downarrow$ & Completion \% $\uparrow$ & Mean horizon \% $\uparrow$ & $I_p$ MAE (kA) $\downarrow$ & Boundary (cm) $\downarrow$ & $\kappa$ $\downarrow$ & Completion \%, noise-trained $\uparrow$ \\
  \midrule
  D1 surrogate & 0.829 & 66.7 & 88.5 & 10.6 & 1.9 & 0.048 & 0.0 \\
  D2 surrogate & 0.708 & 83.3 & 99.0 & 9.3 & 1.4 & 0.063 & 50.0 \\
  D3 surrogate & 0.654 & 83.3 & 94.7 & 11.9 & 1.9 & 0.125 & 16.7 \\
  D4 surrogate & 0.381 & 83.3 & 90.5 & 9.6 & 1.9 & 0.138 & 16.7 \\
  \bottomrule
\end{tabular}

%% file: sections/conclusion.tex
\section{Conclusion}

Only joint-coverage D4 improves cross-condition prediction monotonically, whereas D1 improves in distribution but degrades out of distribution.  D2--D4 share a completion rate yet differ in duration and tracking, and action-noise-trained policies terminate earlier; prediction error alone is therefore insufficient.  Evidence is limited to \fge{}, one policy and evaluation per cell, confounded model conditioning, and no hardware transfer.

%% file: sections/appendix.tex
\section{EXL-50U platform and supporting validation evidence}
\label{app:platform}

\subsection{Device and real-time control interface}
EXL-50U is a compact spherical torus developed for high-current and advanced-configuration studies.  Its operating envelope includes a major radius of 0.6--0.8\,m, minor radius of 0.32--0.50\,m, aspect ratio of approximately 1.4--1.8, toroidal field up to 1.2\,T, and reported plasma current up to 1\,MA \citep{shi2025strategy,shi2025onema,shi2026overview,wsaic2026fusionchallenge}.  The integrated center stack contains the toroidal-field system and central solenoid (CS), while active poloidal-field circuits regulate current, position, and boundary geometry.  The plasma-control system communicates through reflective memory and operates on a 1-ms main cycle.  Plasma current is measured directly; geometric quantities are supplied by fast magnetic reconstruction.  Learned controllers return coil-voltage commands subject to power-supply and machine-protection limits.

\begin{figure}[H]
  \centering
  \includegraphics[width=\linewidth]{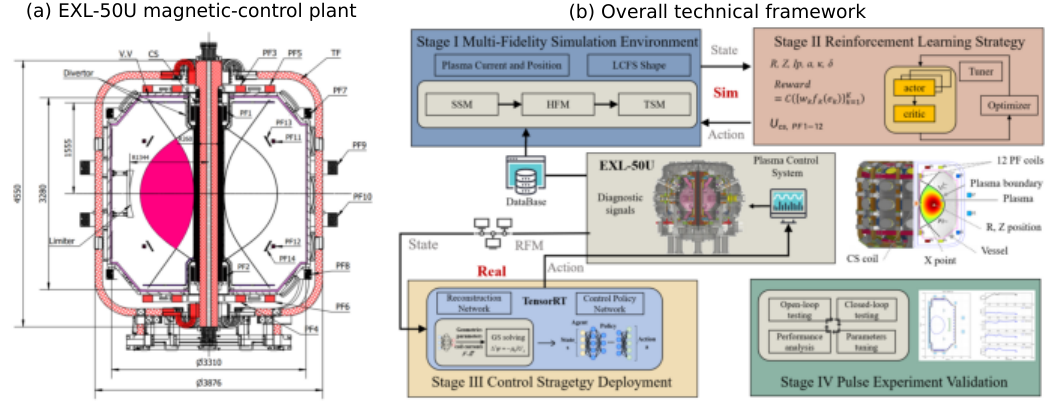}
  \caption{EXL-50U and its overall technical framework.  (a) Original poloidal cross-section showing the central solenoid, toroidal-field system, vessel, limiter, divertor, and poloidal-field coils \citep{guo2026ssmrl}.  (b) Original end-to-end framework covering multi-fidelity simulation, reinforcement-learning training, real-time deployment, and pulse validation \citep{ding2026xpt}.  The present study uses the FGE/TSM modeling path and the $I_p$, $R_{\min}$, $R_{\max}$, and $\kappa$ control interface.}
  \label{fig:platform-stack}
\end{figure}

\begin{table}[H]
  \caption{EXL-50U parameters and interfaces relevant to this study.  Values summarize the published device overviews and the supplied competition specification; they describe the device envelope rather than the operating point of every discharge.}
  \label{tab:device-overview}
  \centering
  \small
  \begin{tabular}{ll}
    \toprule
    Item & Value or role \\
    \midrule
    Major/minor radius & 0.6--0.8\,m / 0.32--0.50\,m \\
    Aspect ratio / elongation & approximately 1.4--1.8 / approximately 2 \\
    Toroidal field / reported peak $I_p$ & up to 1.2\,T / 1\,MA \\
    Main magnetic actuators & CS and active PF circuits; VS handled separately when required \\
    Real-time cycle & 1\,ms (1\,kHz) \\
    Controller interface & magnetic reconstruction and coil currents $\rightarrow$ coil voltages \\
    Protection & voltage/current limits, interlocks, and conventional-controller fallback \\
    \bottomrule
  \end{tabular}
\end{table}

\subsection{Simulator and validation}
FGE is the reference simulator used in this paper.  It evolves a free-boundary Grad--Shafranov equilibrium together with active/passive conductor circuits and a reduced Ohmic-current model \citep{heiss2026fge,ding2026xpt}.  Given PF-coil voltages, the solver advances the poloidal-flux equilibrium, conductor currents, and bulk plasma current, and returns current and boundary descriptors.  The EXL-50U implementation uses a $66\times65$ $(R,Z)$ grid, 544 passive filaments, 12 active circuits, and a fixed 1-ms step; $\beta_p$ and $q_0$ constrain the free profiles, while bulk plasma resistance closes the reduced current evolution.

\begin{figure}[H]
  \centering
  \begin{minipage}[t]{0.62\linewidth}
    \centering
    \includegraphics[width=\linewidth]{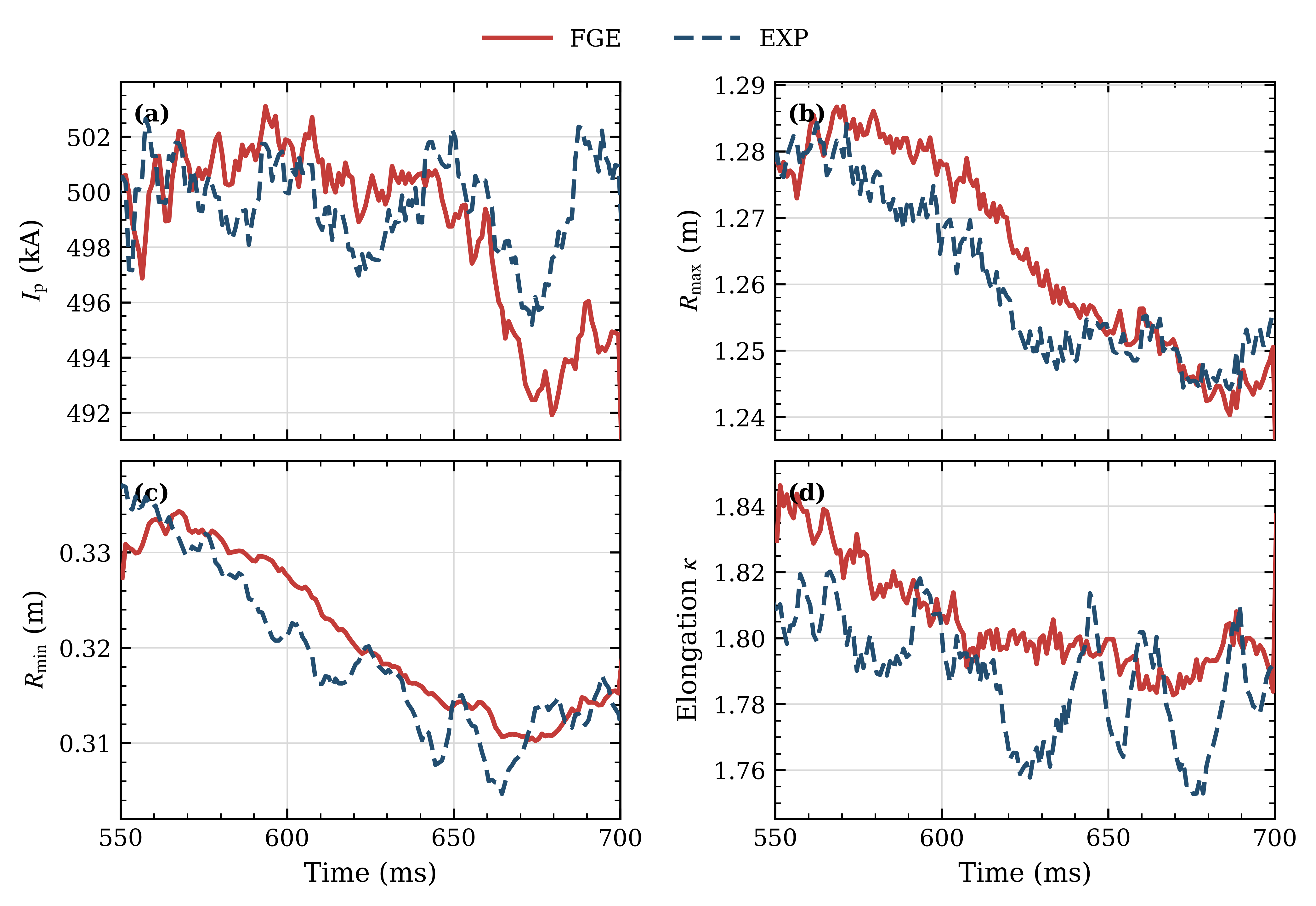}
    \caption{Temporal validation of FGE against experimental reconstruction for EXL-50U discharge \#13844 over 550--700\,ms: $I_p$, $R_{\max}$, $R_{\min}$, and elongation $\kappa$.}
    \label{fig:fge-temporal-validation}
  \end{minipage}\hfill
  \begin{minipage}[t]{0.31\linewidth}
    \centering
    \includegraphics[width=\linewidth]{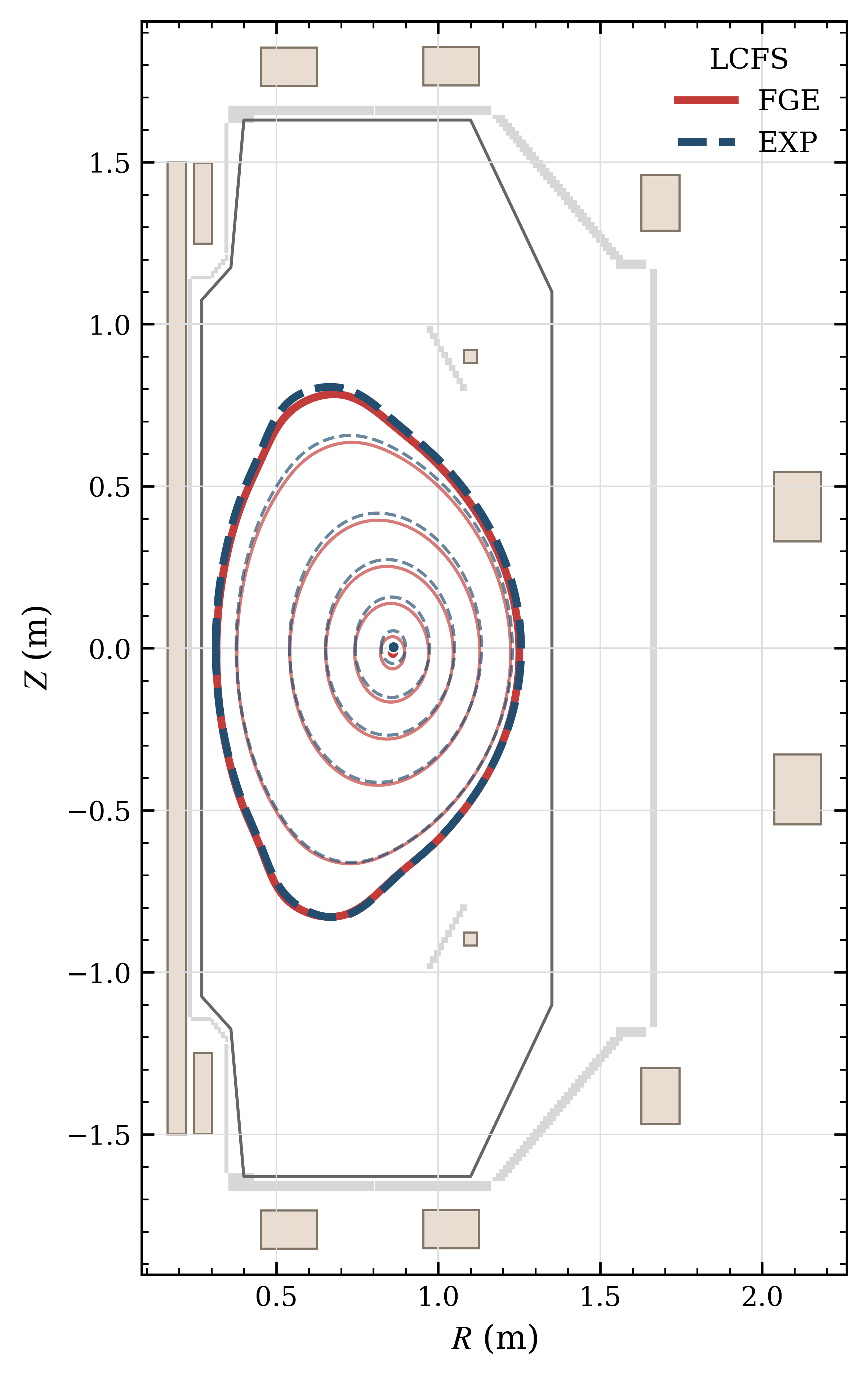}
    \caption{Flux-surface fitting for discharge \#13844 at 700\,ms, comparing FGE with experimental reconstruction.}
    \label{fig:fge-profile-validation}
  \end{minipage}
\end{figure}

For discharge \#13844, FGE is compared with experimental equilibrium reconstruction in both temporal descriptors and flux-surface geometry.  Figures~\ref{fig:fge-temporal-validation} and \ref{fig:fge-profile-validation} show agreement in the principal trends of $I_p$, $R_{\min}$, $R_{\max}$, and $\kappa$, and in the reconstructed last closed flux surface at 700\,ms.  Residual mismatches caution against treating FGE as an exact digital twin, but support its use as a control-oriented magnetic model.

\subsection{On-device trajectory-switching experiment}
\begin{figure}[H]
  \centering
  \includegraphics[width=\linewidth]{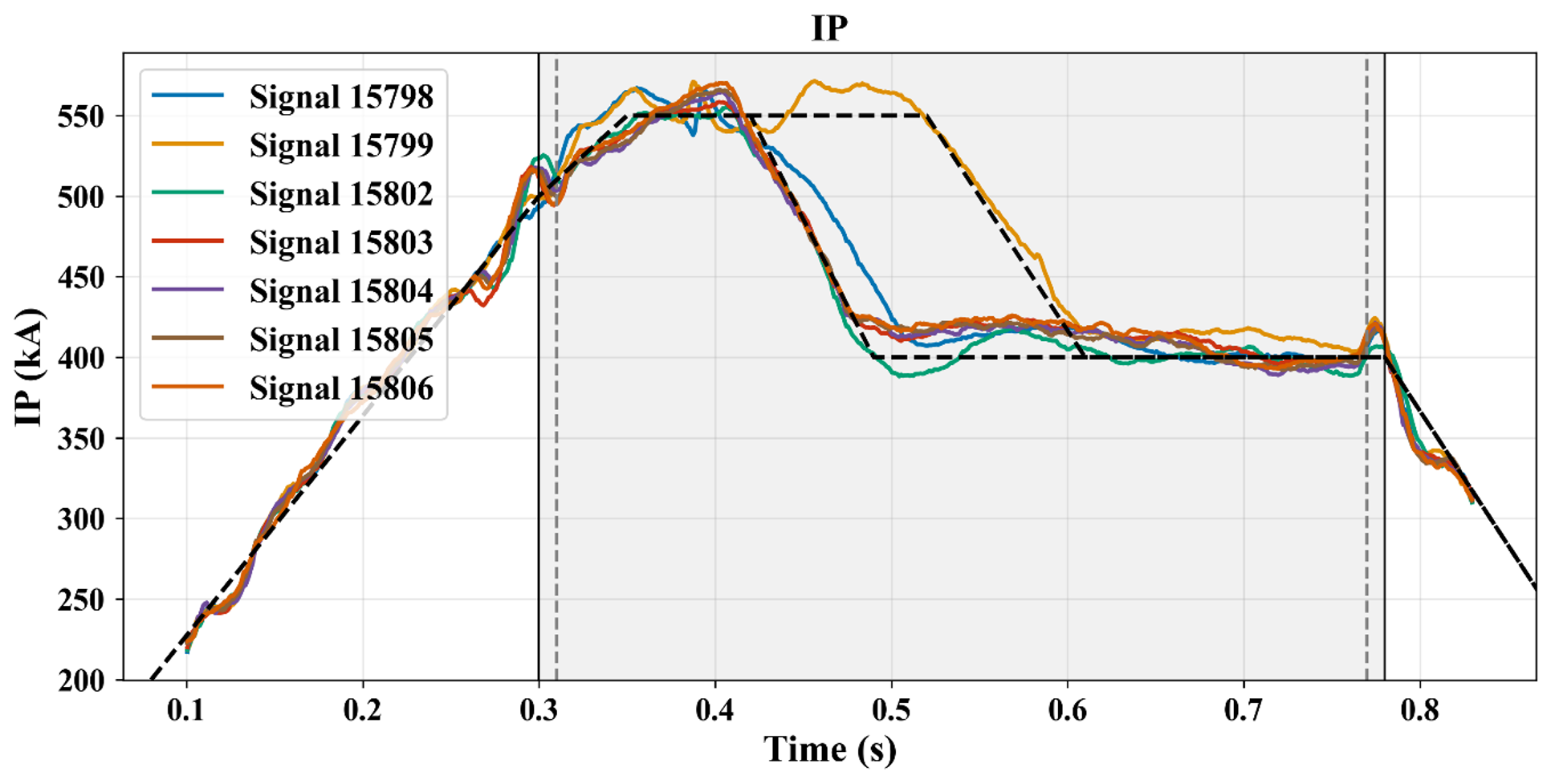}
  \caption{EXL-50U on-device trajectory-switching experiment.  The dashed curve denotes the requested plasma-current trajectory and the colored curves denote seven experimental discharges.}
  \label{fig:fge-device-closed-loop}
\end{figure}

Figure~\ref{fig:fge-device-closed-loop} presents an independently conducted EXL-50U trajectory-switching experiment.  The requested plasma current first rises to approximately 550\,kA, then switches to a 400-kA plateau before the final ramp-down.  Seven experimental discharges reproduce the main target transitions, with shot-to-shot differences concentrated around the switching transients.  This experiment provides independent on-device evidence for time-varying current tracking and is distinct from the D1--D4 reference-solver evaluations in the main study.

\section{Implementation and reproducibility details}
\label{app:impl}

\subsection{Temporal-model state and alignment}
The dynamic state is
\[
  x_t=[I_t^{\mathrm{CS}},I_t^{\mathrm{PS1:10}},I_t^{\mathrm{VS}},
  I_{p,t},R_t,Z_t,R_{\max,t},R_{\min,t},\kappa_t,r_{c,t}]
  \in\mathbb{R}^{19}.
\]
The first 12 entries are circuit currents.  Parameter-conditioned models
add the rollout-constant context
\[
  \phi=(L_p^{\mathrm{scale}},b_p,q_0,\sigma_{\mathrm{neo}})
  \in\mathbb{R}^{4},
\]
whereas base models omit it.  The action $a_t\in\mathbb{R}^{12}$ contains the
corresponding \texttt{CS\_UOUT}, \texttt{PS1\_UOUT}--\texttt{PS10\_UOUT},
and \texttt{VS\_UOUT} voltages.

Each stored transition follows the simulator execution order
\[
  x_{t-1},a_t \longmapsto x_t.
\]
Consequently, a context ending at $x_t$ is paired with the action ending at
$a_{t+1}$ when predicting $x_{t+1}$.  State and action channels are
standardized independently using training-set means and standard deviations,
and the stored statistics are reused at validation and rollout time.  The next
circuit currents are predicted rather than copied from the recorded future,
preventing future-state leakage.  At each online step, the oldest
state--action pair is discarded and the newest command and prediction are
appended; the control interface reads $(I_p,R_{\min},R_{\max},\kappa)$.

\subsection{Transformer and multi-step training}
Each token concatenates one dynamic state, its aligned action, and, for D2/D4,
the broadcast physics context.  The input width is therefore 31 for the base
model and 35 for the parameter-conditioned model.  All configurations use a
three-block Transformer with $d_{\mathrm{model}}=128$, eight attention heads,
a 512-dimensional feed-forward sublayer, dropout 0.1, and non-causal
self-attention over the observed context.  The archived configuration enables
input padding and uses a history of $H=30$ steps.  The selected D4
configuration lists 23 state/context channels; during evaluation, the four
static physics values are held fixed and excluded from the reported dynamic
prediction error.

Training unrolls the one-step model for $K=10$ steps.  In standardized
coordinates, the objective is
\begin{align}
  \hat{x}_{t+k}
    &= f_\theta(\hat{x}_{t+k-H:t+k-1},
       a_{t+k-H+1:t+k},\phi),\\
  \mathcal{L}_{\mathrm{AR}}
    &= \frac{1}{K}\sum_{k=1}^{K}
       \left\|\hat{x}_{t+k}-x_{t+k}\right\|_2^2 .
\end{align}
After each prediction, the oldest state is removed and $\hat{x}_{t+k}$ is fed
back, so steps 2--10 are optimized under model-generated state context.
Base-model equations omit $\phi$.

\begin{table}[H]
  \caption{Temporal-surrogate training settings recorded in the archived
  configuration.}
  \label{tab:tsm-hparams}
  \centering
  \small
  \begin{tabular}{lc@{\qquad}lc}
    \toprule
    Setting & Value & Setting & Value \\
    \midrule
    History / unroll & 30 / 10 & Model / FF width & 128 / 512 \\
    Attention heads / blocks & 8 / 3 & Dropout & 0.1 \\
    Batch size & 128 & Window stride & 10 \\
    Learning rate & $10^{-4}$ & Gradient-norm clip & 1 \\
    Optimizer / epochs & Adam / 20 & Seed & 42 \\
    \bottomrule
  \end{tabular}
\end{table}

Windows are sampled with stride 10, but partitioning occurs at the complete
rollout level.  The archived selected D4 configuration records a
$10/11$--$1/11$ training--validation split and no internal test allocation;
C1--C4 are supplied through separate held-out manifests.  Validation uses
recursive multi-step loss, and the checkpoint with the lowest validation loss
is retained.

\subsection{Collection distributions}
Nominal plasma parameters are $b_p=0.21$, $q_0=1.5$,
$\sigma_{\mathrm{neo}}=7.5\times10^6$, and a multiplicative $L_p$ scale of 1; the randomized
configurations scale them over $[0.70,1.30]$, $[0.80,1.30]$, $[0.70,1.30]$, and
$[0.80,1.00]$ respectively.  The seven
independent policy actions correspond to CS, paired PS1/2, PS3/4, PS5/6,
PS7/8, PS9/10, and VS commands.  During collection, temporally correlated
colored action noise is added before clipping and rate limiting,
using per-channel voltage scales $(250,40,40,40,60,60,50)$.  The surrogate is
fitted on the 12 voltages actually applied after both constraints.

\begin{figure}[H]
  \centering
  \includegraphics[width=\linewidth]{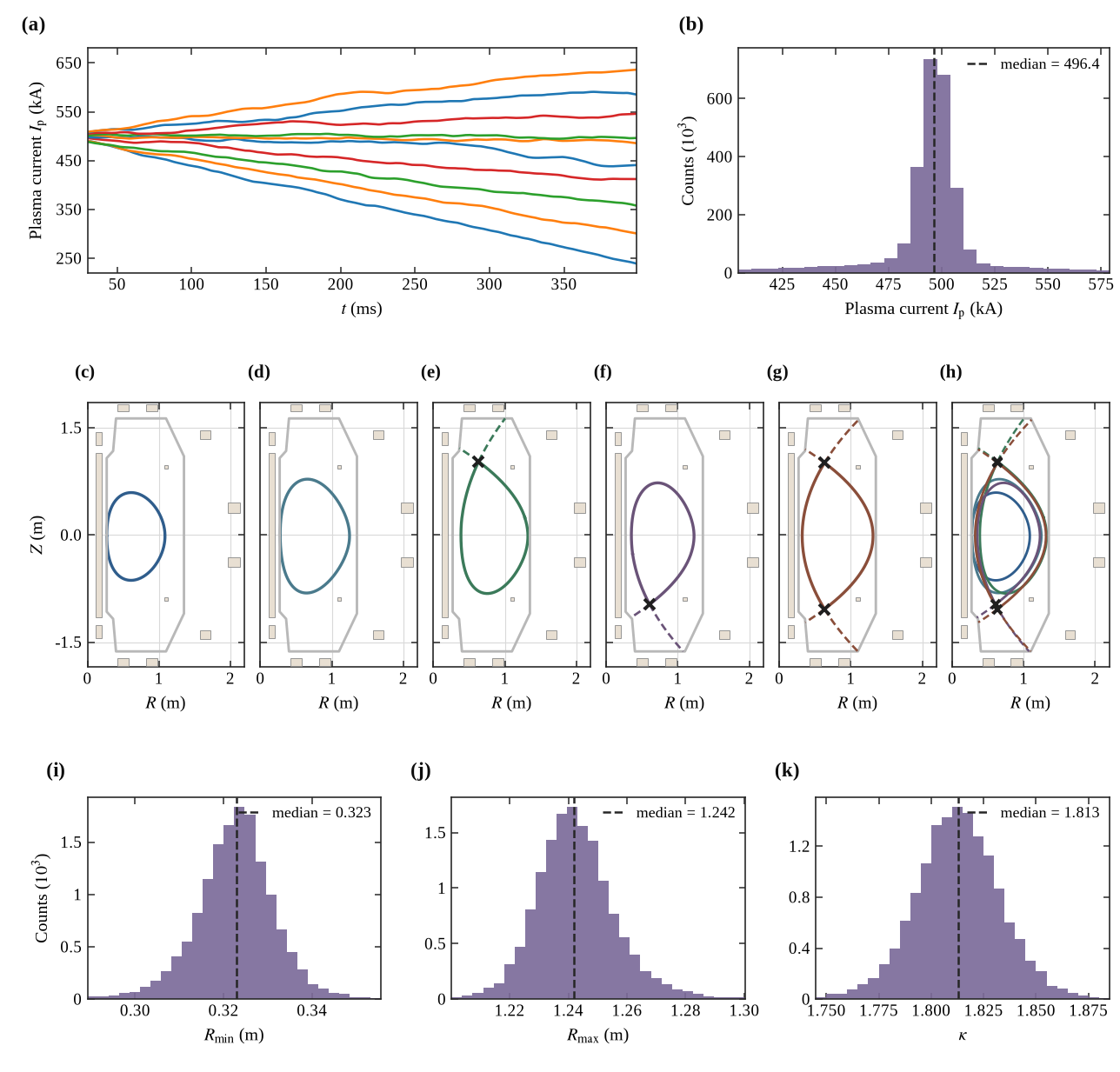}
  \caption{Approximate distribution of the collected data.  (a) Representative plasma-current trajectories and (b) their marginal distribution.  (c)--(h) Representative plasma-boundary configurations.  (i)--(k) Marginal distributions of $R_{\min}$, $R_{\max}$, and elongation $\kappa$.}
  \label{fig:data-distribution}
\end{figure}

\subsection{Policy training interface}
PPO is trained separately in each frozen surrogate.  The policy receives the
normalized tracking variables $(I_p,R_{\min},R_{\max},\kappa)$ and emits seven
commands in $[-1,1]$.  They are rescaled to voltage lower bounds
$(-1499,-230,-172,-172,-348,-348,-270)$ and upper bounds
$(100,230,172,172,348,348,270)$\,V.  The fixed map
$(0,1,1,2,2,3,3,4,4,5,5,6)$ expands these commands to CS, five paired PF
groups, and VS, yielding the 12 voltage inputs consumed by the surrogate.
Deterministic policy actions are used during evaluation.

For each D1--D4 surrogate, the nominal and noise-trained runs share the PPO
network, optimizer settings, training budget, reward, action bounds, and
termination logic; only the training-time action perturbation is changed.
This perturbation is distinct from the colored excitation used to collect the
surrogate dataset.  One policy checkpoint is retained per
(surrogate, training-regime) pair.  The evaluated H5 files identify these
controllers as Ray-based trainers.  Because they do not embed the complete
trainer configuration, backend-specific quantities such as minibatch size,
GAE parameters, clipping coefficient, and policy seed cannot be independently
reconstructed from the paper artifact and are not inferred from the separate
SB3 example configuration.

\section{Evaluation protocol}

\paragraph{Closed-loop evaluation.}
Every frozen controller runs in \fge{} at \ControlHz{}\,Hz for \EvalSteps{}
steps (\EvalSeconds{}\,s), ramping $I_p$ from 495\,kA to one of 400/500/600\,kA
while holding the boundary shape.  The two test environments are the ramp at
nominal physics and the ramp with the physics parameters additionally
perturbed.  Controllers reuse the same random seeds, preserving randomized
parameter samples across policies, so the \TrialsPerCell{} conditions for each
(policy, training regime) combination form a paired evaluation.  Reward, action
bounds, and termination conditions are frozen before ranking.

\paragraph{Episode completion and tracking metrics.}
An episode is counted as complete only if it reaches all \EvalSteps{} steps;
any earlier \fge{} termination counts as failure, as determined from trajectory
length.  Episode length is capped at \EvalSteps{} steps, so duration summaries
use $\min(T,\EvalSteps)$.  A shorter failed episode may yield a deceptively
small time-averaged tracking error.  We therefore report tracking error among
completed episodes only, together with the number of episodes included.  In the
displayed physical units, channel-wise MAE is
\[
  E_j(\pi)=\frac{1}{T}\sum_{t=1}^{T}
  |s_{t,j}-s^\star_{t,j}|,
\]
after fixed conversion to $I_p$ in kA, boundary radii in cm, and dimensionless
$\kappa$.  These quantities are reported separately and never summed.
Figure~\ref{fig:results}b uses a separate set of direct TSM closed-loop tests:
one 200-step trajectory per D1--D4 surrogate at a fixed 495-kA target, using
the policy trained without action noise.  These surrogate-internal traces are
not included in the six-condition FGE completion and tracking aggregates.

\paragraph{Open-loop evaluation.}
The predictive test set comprises held-out trajectories from the frozen RL
collector; no training windows are extracted from these trajectories.
Autoregressive error applies the channel-wise normalization used in
preprocessing before averaging.  The equal-weight average over C1--C4 is the
cross-condition mean error used throughout.

\section{Complete result tables}
\label{app:tables}

The open-loop table reports all 16 models across the four matched budgets; the closed-loop table reports all eight policies trained in the selected 5k surrogates.  All values are generated by \path{scripts/build_results.py} from \path{data/openloop_results.csv} and \path{data/control_results.csv}.

\begin{table}[H]
  \caption{Autoregressive error for the \NumTemporalModels{} temporal models on all four held-out conditions.  Each dataset contributes one model at each rollout budget: D1 and D3 use the base version, whereas D2 and D4 use the parameter-conditioned version.  The last column is the equal-weight cross-condition average; diagonal entries are in-distribution errors.}
  \label{tab:app-openloop}
  \centering
  \scriptsize
  \input{tables/app_openloop_en}
\end{table}

\begin{figure}[H]
  \centering
  \includegraphics[width=0.72\linewidth]{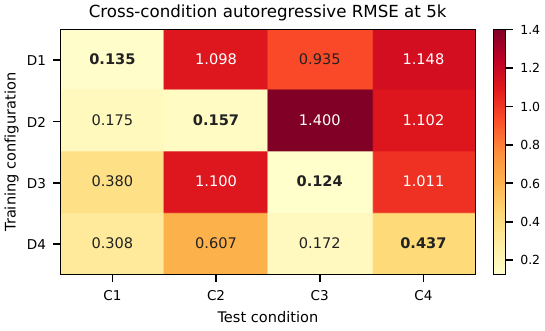}
  \caption{The 5k cross-condition error matrix.  Bold entries are the lowest error in each test condition.  The lowest error in each column is attained by the configuration matched to that condition, illustrating why in-distribution errors are not directly comparable across configurations.}
  \label{fig:app-openloop-matrix}
\end{figure}

\begin{table}[H]
  \caption{Per-environment closed-loop results for every policy.  ``Completed'' counts episodes reaching \EvalSteps{} steps out of those attempted (three current endpoints per environment); tracking error covers completed episodes only and is blank where none complete.  Steady $I_p$ is the mean absolute error over the final quarter of the trajectory.}
  \label{tab:app-control}
  \centering
  \scriptsize
  \input{tables/app_control_en}
\end{table}

\begin{table}[H]
  \caption{Condition-level duration for all 48 trials.  A check mark denotes completion of all \EvalSteps{} steps; numbers give the termination step.  The six columns are distinct test conditions, not repeated random trials.}
  \label{tab:app-steps}
  \centering
  \scriptsize
  \input{tables/app_steps_en}
\end{table}

\begin{table}[H]
  \caption{Tracking error on the \CommonSurvivorPairs{} matched conditions completed by both nominal- and action-noise-training policies.  Restricting the comparison to the same completed conditions avoids comparing different test subsets but does not remove single-policy uncertainty.}
  \label{tab:app-paired-tracking}
  \centering
  \scriptsize
  \input{tables/app_paired_tracking_en}
\end{table}

\begin{figure}[H]
  \centering
  \includegraphics[width=\linewidth]{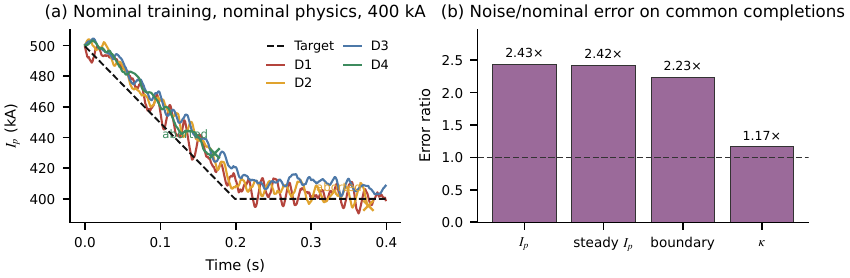}
  \caption{Closed-loop diagnostics.  (a) Representative 400\,kA trajectories under nominal training and nominal physics; crosses mark early termination.  (b) Action-noise to nominal error ratios on the \CommonSurvivorPairs{} matched conditions completed under both regimes; values above one indicate larger error under action-noise training.}
  \label{fig:app-control-diagnostics}
\end{figure}

\section{Limitations}

\paragraph{External validity.}
The controlled datasets originate from \fge{}, not directly from diverse
experimental discharges.  The results therefore characterize
surrogate-to-reference policy performance, not hardware transfer.

\paragraph{Coverage and information volume.}
All four datasets use one frozen RL collector with the same excitation distribution,
each contains 5k rollouts, and every configuration is retrained at 500/1k/3k/5k.
Coverage is therefore compared at matched trajectory counts.
The remaining difference concerns information rather than count: a broad-coverage rollout carries
more diverse dynamics than a narrow one, so matched counts are not matched
effective samples.  Isolating the causal effect of coverage would require both
matching effective sample diversity or state--action occupancy coverage and
holding model conditioning fixed.  D1 and D3 use the base model, while D2 and D4 use the
same parameter-conditioned model corresponding to their physics-coverage
configurations.

\paragraph{Statistical precision and mechanism.}
Each configuration contains \TrialsPerCell{} distinct paired conditions, not repeated
trials, with one evaluation realization per condition and one trained PPO policy.
Completion rates are therefore
descriptive, with resolution $1/\TrialsPerCell{}$; no confidence intervals or
fine-grained ordering among nearby cells are claimed.  The termination step
identifies failure but does not distinguish current-tracking divergence,
boundary instability, and actuator saturation.  Consequently, the experiments
demonstrate a forecasting--control discrepancy but do not localize its dynamical
mechanism.

%% file: tables/app_openloop_en.tex
\setlength{\tabcolsep}{4pt}
\begin{tabular}{lllccccc}
  \toprule
  Dataset & Episodes & Model variant & C1 nom./fixed & C2 rand./fixed & C3 nom./ramp & C4 rand./ramp & Cross-cond.\ mean \\
  \midrule
  D1 & 500 & base & 0.345 & 0.892 & 0.909 & 1.078 & 0.806 \\
   & 1k & base & 0.240 & 0.953 & 0.833 & 1.058 & 0.771 \\
   & 3k & base & 0.208 & 1.016 & 0.883 & 1.118 & 0.806 \\
   & 5k & base & 0.135 & 1.098 & 0.935 & 1.148 & 0.829 \\
  \midrule
  D2 & 500 & conditioned & 0.271 & 0.221 & 2.468 & 1.919 & 1.220 \\
   & 1k & conditioned & 0.182 & 0.200 & 1.855 & 1.394 & 0.908 \\
   & 3k & conditioned & 0.189 & 0.193 & 2.424 & 1.680 & 1.122 \\
   & 5k & conditioned & 0.175 & 0.157 & 1.400 & 1.102 & 0.708 \\
  \midrule
  D3 & 500 & base & 0.527 & 0.979 & 0.298 & 0.832 & 0.659 \\
   & 1k & base & 0.436 & 1.076 & 0.261 & 0.904 & 0.669 \\
   & 3k & base & 0.402 & 1.159 & 0.227 & 0.984 & 0.693 \\
   & 5k & base & 0.380 & 1.100 & 0.124 & 1.011 & 0.654 \\
  \midrule
  D4 & 500 & conditioned & 0.571 & 0.741 & 0.305 & 0.497 & 0.528 \\
   & 1k & conditioned & 0.463 & 0.655 & 0.298 & 0.474 & 0.473 \\
   & 3k & conditioned & 0.370 & 0.573 & 0.185 & 0.430 & 0.389 \\
   & 5k & conditioned & 0.308 & 0.607 & 0.172 & 0.437 & 0.381 \\
  \bottomrule
\end{tabular}

%% file: tables/app_control_en.tex
\setlength{\tabcolsep}{4pt}
\begin{tabular}{lllccccc}
  \toprule
  Policy & Training & Env. & Completed & $I_p$ MAE & Steady $I_p$ & Boundary (cm) & $\kappa$ \\
  \midrule
  D1 surrogate & nominal & nominal physics & 1/3 & 6.8 & 3.8 & 1.5 & 0.014 \\
   &  & perturbed physics & 3/3 & 11.9 & 12.0 & 2.1 & 0.060 \\
   & action noise & nominal physics & 0/3 & -- & -- & -- & -- \\
   &  & perturbed physics & 0/3 & -- & -- & -- & -- \\
  \midrule
  D2 surrogate & nominal & nominal physics & 2/3 & 10.9 & 17.7 & 1.6 & 0.034 \\
   &  & perturbed physics & 3/3 & 8.3 & 8.3 & 1.2 & 0.082 \\
   & action noise & nominal physics & 0/3 & -- & -- & -- & -- \\
   &  & perturbed physics & 3/3 & 27.2 & 20.2 & 2.7 & 0.135 \\
  \midrule
  D3 surrogate & nominal & nominal physics & 2/3 & 9.0 & 5.6 & 1.7 & 0.044 \\
   &  & perturbed physics & 3/3 & 13.9 & 11.5 & 2.0 & 0.179 \\
   & action noise & nominal physics & 0/3 & -- & -- & -- & -- \\
   &  & perturbed physics & 1/3 & 27.4 & 38.9 & 3.9 & 0.196 \\
  \midrule
  D4 surrogate & nominal & nominal physics & 2/3 & 5.0 & 3.2 & 1.4 & 0.087 \\
   &  & perturbed physics & 3/3 & 12.7 & 10.4 & 2.2 & 0.171 \\
   & action noise & nominal physics & 0/3 & -- & -- & -- & -- \\
   &  & perturbed physics & 1/3 & 16.6 & 11.6 & 4.2 & 0.080 \\
  \bottomrule
\end{tabular}

%% file: tables/app_steps_en.tex
\begin{tabular}{llcccccc}
  \toprule
  Policy & Training & Nom. 400 & Nom. 500 & Nom. 600 & Pert. 400 & Pert. 500 & Pert. 600 \\
  \midrule
  D1 & nominal & \checkmark & 374 & 150 & \checkmark & \checkmark & \checkmark \\
   & action noise & 31 & 30 & 33 & 24 & 27 & 31 \\
  \midrule
  D2 & nominal & 377 & \checkmark & \checkmark & \checkmark & \checkmark & \checkmark \\
   & action noise & 262 & 326 & 191 & \checkmark & \checkmark & \checkmark \\
  \midrule
  D3 & nominal & \checkmark & \checkmark & 273 & \checkmark & \checkmark & \checkmark \\
   & action noise & 117 & 149 & 300 & 128 & \checkmark & 240 \\
  \midrule
  D4 & nominal & 173 & \checkmark & \checkmark & \checkmark & \checkmark & \checkmark \\
   & action noise & 145 & 170 & 196 & 205 & 375 & \checkmark \\
  \bottomrule
\end{tabular}

%% file: tables/app_paired_tracking_en.tex
\begin{tabular}{lccccc}
  \toprule
  Training & $n$ & $I_p$ MAE (kA) & Steady $I_p$ & Boundary (cm) & $\kappa$ \\
  \midrule
  nominal & 5 & 10.3 & 9.2 & 1.5 & 0.117 \\
  action noise & 5 & 25.1 & 22.2 & 3.3 & 0.136 \\
  \bottomrule
\end{tabular}